\documentclass[12pt]{iopart}
\usepackage{iopams}
\begin{document}
\newcommand{\beq}{\begin{equation}} 
\newcommand{\eeq}{\end{equation}} 
\newcommand{\bea}{\begin{eqnarray}} 
\newcommand{\eea}{\end{eqnarray}} 
\newcommand{\bear}{\begin{array}}
\newcommand{\eear}{\end{array}} 
\newcommand{\LM}{{\bf L}} 
\newcommand{\VM}{\bf V } 
\newcommand{\N}{{\bf N}} 
\newcommand{\la}{{\lambda}} 
\newcommand{\larra}{{\longrightarrow}} 
\newcommand{\TLM}{{\bf \tilde{N}}} 
\newcommand{\TQ}{{\tilde{Q}}} 
\newcommand{\TP}{{\tilde{P}}} 
\newcommand{\UQ}{{\underline{Q}}} 
\newcommand{\dt}{\mathrm{det}} 
\newcommand{\ovl}{\overline} 
\newcommand{\unl}{\underline} 
\newcommand{\lm}{\raisebox{-1.ex}{\em -}} 
\newcommand{\lp}{\raisebox{-1.ex}{\em +}}
\newtheorem{Pro}{Proposition}[section]
\newtheorem{Thm}{Theorem}[section]
\newtheorem{Lem}{Lemma}[section]
\newtheorem{Cor}{Corollary}[section] 
\newtheorem{Def}{Definition}[section]  
\newcommand{\hp}{\hat{p}} 
\newcommand{\hq}{\hat{q}}  
\newcommand{\ta}{\tau} 

\newenvironment{prf}{\trivlist \item [\hskip
\labelsep {\bf Proof:}]\ignorespaces}{ \endtrivlist}


\title[Rational solutions of discrete Toda and $alt$-$dP_{II}$] 
{Rational solutions of the discrete time Toda lattice and the 
alternate discrete 
Painlev\'{e}  II equation} 

\author{Alan K. Common and Andrew N. W. Hone\dag} 


\address{\dag\ Institute of Mathematics, Statistics \&
Actuarial Science,
University of Kent,
Canterbury CT2~7NF, UK}

\ead{A.N.W.Hone@kent.ac.uk}

\begin{abstract} 
The Yablonskii-Vorob'ev polynomials $y_{n}(t)$, which are defined by 
a second order bilinear 
differential-difference equation, provide rational solutions 
of the Toda lattice. 
They are also polynomial 
tau-functions for the rational solutions of the second 
Painlev\'{e} equation ($P_{II}$). Here we define two-variable  
polynomials $Y_{n}(t,h)$ on a lattice with spacing $h$, 
by considering rational solutions of the discrete time Toda lattice 
as introduced by Suris. 
These polynomials are shown to have many properties that are 
analogous to those of 
the Yablonskii-Vorob'ev polynomials, 
to which they reduce when $h=0$. 
They also 
provide rational solutions for a particular discretisation of $P_{II}$, 
namely the so called {\it alternate discrete} $P_{II}$, 
and this connection leads to an expression in terms of the Umemura 
polynomials for the third Painlev\'{e} equation ($P_{III}$). 
It is shown that 
B\"{a}cklund transformation for the alternate discrete Painlev\'{e} equation 
is a symplectic map, 
and the shift in time is also symplectic. Finally we  
present a Lax pair for the alternate discrete $P_{II}$,
which recovers Jimbo and Miwa's Lax pair for 
$P_{II}$ in the continuum limit $h\to 0$. 
\end{abstract}

\submitto{J. Phys. A: Math. Theor.}

\maketitle
 
\newpage 
\section{Introduction}

The Toda lattice 
\beq 
\frac{d^2x_n}{dt^2}=e^{x_{n-1}-x_n}-e^{x_n-x_{n+1}}, \qquad n\in\mathbb{Z} 
\label{toda} 
\eeq  
was the first integrable differential-difference 
equation to be discovered \cite{tod1}.  
The Yablonskii-Vorob'ev polynomials \cite{Yab, Vob} 
yield rational solutions of both the Toda lattice and     
the second Painlev\'{e} transcendent ($P_{II}$), since 
the tau-functions of $P_{II}$ satisfy the bilinear form of the Toda lattice. 
In a previous work \cite{haf1} one of the authors 
obtained an expression for solutions of the Toda lattice as 
ratios of Hankel determinants, by using the associated Lax pair 
to construct continued fraction solutions to a sequence of Riccati 
 equations. This in turn led to an expression for the 
Yablonskii-Vorob'ev poynomials as Hankel determinants \cite{haf2}, 
equivalent 
to that discovered more recently \cite{KO} (see also \cite{josh}).
 
Here we will consider the case when the time evolution becomes discrete. 
Our approach is to start from the Lax pair for the discrete Toda 
lattice (dTL) given by Suris \cite{sur}. 
In section 2 we present this construction and derive polynomials 
$Y_{n}(t,h)$ in two variables $t,h$, which satisfy  
relations at a discrete set of points $ t=t_0+mh$, $m\in\mathbb{Z}$ 
(where  $t_0$ is arbitrary).  
These polynomials tend to the Yablonskii-Vorob'ev polynomials $y_{n}(t)$ as 
the spacing $h\rightarrow0$, so that $Y_n (t,0)=y_n (t)$. 
A discrete analogue of the bilinear defining equation for the 
$y_{n}$ 
is derived and a corresponding representation of these $Y_n$  
as Hankel determinants is given.

The first few Yablonskii-Vorob'ev polynomials are 
\beq \label{pols}
\fl 
y_0=1,\,\, y_1=t, \, \, y_2=t^3+4, \,\, y_3=t^6+20t^3-80,\,  
\, y_4=t^{10}+60t^7+11200t. 
\eeq 
A further property of these 
polynomials, 
investigated recently by Clarkson and Mansfield \cite{pet2}, 
is the distribution of their zeroes. It is known that 
each $y_n(t)$ has no zeroes in common with $y_{n\pm 1}(t)$, and that 
these zeroes are simple \cite{ok, tan}. 
Numerical studies indicate that these zeroes lie in approximately triangular arrays 
and that the zeroes of $y_{n}(t)$ 
interlace, 
in a certain sense, with those of $y_{n+1}(t)$, in a similar way 
to the zeroes 
of classical orthogonal polynomials. In section 3 we show that 
the polynomials $Y_{n}(t,h)$ have analogous properties. 

In the continuum case the bilinear differential-difference equation that 
defines the Yablonskii-Vorob'ev polynomials 
is given in terms of the Hirota derivative $D_t$ by 
\beq \label{biltoda} 
y_{n+1}y_{n-1}=ty_{n}^2-2D_t^2 y_n\cdot y_n, 
\eeq   
with the initial polynomials $y_0=1$, $y_1=t$, 
and this equation follows from the representation of $P_{II}$ as  
a pair of bilinear equations for the associated tau-functions. 
In section 4 we consider a discretisation of $P_{II}$, 
namely the alternate discrete $P_{II}$ equation 
($alt$-$dP_{II}$) 
studied in \cite{FSKGR}, 
and explain how its solutions are 
specified by tau-functions which satisfy a 
discrete bilinear equation together with a 
quadrilinear (degree four) relation. 
The bilinear relation defining the polynomials  
$Y_{n}(t,h)$ is a consequence of these tau-function equations.  
The $alt$-$dP_{II}$ equation has B\"{a}cklund transformations 
corresponding to the shifts $n \to n \pm 1$, just as in 
the continuum case, which imply that  
the discrete Yablonskii-Vorob'ev polynomials  
satisfy recurrence relations which are analogous 
to those in the continuum setting. It is also known that 
the $alt$-$dP_{II}$ equation arises as the contiguity relations 
for a B\"{a}cklund transformation for $P_{III}$. 
The latter connection leads to an alternative formula for the 
discrete Yablonskii-Vorob'ev polynomials
in terms of determinants of Jacobi-Trudi type, 
corresponding to Umemura polynomials \cite{KM}.

It is known from the work of Okamoto that the continuum 
$P_{II}$ can be written in Hamiltonian form, 
and   
the B\"{a}cklund transformation is a canonical transformation \cite{okamoto}.  
In section 5 we present a Poisson structure for 
the discrete case such that both development
in time and the B\"{a}cklund transformation for 
$alt$-$dP_{II}$ are 
symplectic maps.      
Finally in section 6 we present a Lax pair 
for $alt$-$dP_{II}$, which relates it to 
the isomonodromic deformation of an associated 
linear system, and show that this tends to Jimbo 
and Miwa's Lax pair for the 
continuum case as $h\rightarrow 0$.  
 
\section{Discrete time Toda lattice and discrete Yablonskii-Vorob'ev 
polynomials} 
    
The equations for the discrete Toda lattice 
obtained by Suris \cite{sur} are   
\beq 
\bear{ccl} 
x_{n}(t+h)-x_{n}(t)&=&h \pi_{n}(t+h), \\ 
\pi_{n}(t+h)-\pi_{n}(t)&=&\frac{1}{h}\log 
\left[\frac{1+h^2\exp(x_{n-1}(t)-x_{n}(t))}{1+h^2\exp(x_{n}(t)-x_{n+1}(t))}\right] ,

\label{eq:dtl1}

\eear    
\eeq  
where $\pi_n$ denotes the canonically conjugate momentum to 
$x_n$, and these discrete equations clearly yield Hamilton's equations 
for the Toda lattice (\ref{toda}) in the continuum limit, as $h\rightarrow 0$. 
They arise from the consistency condition for the Lax pair     
\beq
\Psi_{n}(t+h)=\mathbf{V}_{n}(t)\Psi_{n}(t), 
\quad  \Psi_{n+1}(t)=\mathbf{L}_{n}(t)\Psi_{n}(t), 
\label{eq:dlax1}
\eeq
that is 
$\mathbf{L}_{n}(t+h)\mathbf{V}_{n}(t)=\mathbf{V}_{n+1}(t)\mathbf{L}_{n}(t)$,
where
\bea
\mathbf{L}_{n}(t)=\left(\begin{array}{cc}
\lambda \exp(-h\pi_{n}(t)) -\frac{1}{\lambda}&h\exp(-x_{n}(t))\\
-h\exp(x_{n}(t)-h\pi_{n}(t))&0
\end{array} \right)
\label{eq:dlax3}
\eea
and
\bea
\mathbf{V}_{n}(t)=\left( \begin{array}{cc}
\frac{1}{\lambda}&-h\exp(-x_{n}(t))\\
h\exp(x_{n-1}(t))&\lambda
\end{array} \right).
\label{eq:dlax4}
\eea 

Upon setting $ \Psi_{n}(t)=[X_{n}(t),Y_{n}(t)]^{T}$ and $Z_{n}(t)=X_{n}(t)Y_{n}(t)^{-1}$, 
we find from the second of (\ref{eq:dlax1}) that
\beq
Z_{n}(t)=-\frac{h\exp(-x_{n}(t))}{\lambda \exp(-h\pi_{n}(t))-\frac{1}{\lambda}+h\exp(-h\pi_{n}(t)+x_{n}(t))Z_{n+1}(t)}.
\label{eq:dcont1}
\eeq 
This recurrence relation can be used to generate the continued fraction expansion 
\beq
\lambda h Z_{0}(t)=\frac{\gamma f_{1}}{1+\gamma g_{1}} \lp \frac{\gamma f_{2}}{1+\gamma g_{2}} \lp \frac{\gamma f_{3}}{1+\gamma g_{3}} \lp \ldots ,
\label{eq:dcont2}
\eeq
where $ \gamma=\lambda^2,f_{1}=h^2\exp(-x_{0}(t)),g_{1}=-\exp(-h\pi_{0}(t))$ and for $n=2,3,\ldots$ we have 
\bea
f_{n}&=&-h^{2}\exp(-x_{n-1}(t)+x_{n-2}(t)-h\pi_{n-2}(t)), \nonumber \\
g_{n}&=&-\exp(-h\pi_{n-1}(t)).
\label{eq:dcont3}
\eea
Similarly from the first of (\ref{eq:dlax1}) we have the discrete time Riccati equation
\beq
\fl 
Z_{0}(t+h)-Z_{0}(t)=-\frac{h}{\lambda} 
e^{-x_{0}(t)}+ 
\left(\frac{1}{\lambda^{2}}-1\right) 
Z_{0}(t)-\frac{h}{\lambda}e^{x_{-1}(t)}Z_{0}(t+h)Z_{0}(t).
\label{eq:dric1}
\eeq
In our previous work \cite{haf3} we took $x_{-1}(t)\to -\infty$ 
so that this became a linear equation. Here we take
$x_{-1}(t)=0$ and set $Z_{0}(t)=\frac{1}{h\lambda}W(t)$ so that 
with $\gamma=\lambda^2$ we find  
\beq 
W(t+h)=-h^2e^{-x_{0}(t)}+\frac{1}{\gamma}W(t) 
-\frac{1}{\gamma}W(t+h)W(t).
\label{eq:dric2}
\eeq

Now the continued fraction (\ref{eq:dcont2}) has expansions both in positive and negative powers of $\gamma$, 
namely  
$W(t)=\sum_{n=1}^{\infty}\alpha_{n}(t)\gamma^{n}$ 
and
$W(t)=\sum_{-\infty}^{n=0}\alpha_{n}(t)\gamma^{n}$ respectively.
Substituting these expansions in turn into (\ref{eq:dric2}) and equating coefficients of corresponding powers of $\gamma$,
we obtain the recurrence relations for their coefficients, given respectively by  
\bea
\alpha_{j+1}(t)&=&\alpha_{j}(t+h)+\sum_{k=1}^{j}\alpha_{k}(t+h)\alpha_{j-k+1}(t), \quad j=1,2,3,\ldots , \\
\alpha_{j}(t+h)&=&\alpha_{j+1}(t)-\sum_{k=j+1}^{0}\alpha_{k}(t+h)\alpha_{j+1-k}(t), \quad j=-1,-2,\ldots .
\label{eq:dhank1}
\eea
where $\alpha_{1}(t)=h^2\exp(-x_{0}(t))$ and $\alpha_{0}(t+h)=-h^2\exp(-x_{0}(t))$.

The continued fraction in (\ref{eq:dcont2}) is known as a \emph{T-fraction} \cite{jones} and its elements are given
in terms of Hankel determinants. If we take the definitions
\beq
u_{n}=H_{n}^{(-n+2)}, \qquad v_{n}=H_{n}^{(-n+1)},
\label{eq:dhank2}
\eeq
with Hankel determinants
\bea
H_{k}^{(m)}=\left| \begin{array}{cccc}
\beta_{m}&\beta_{m+1}&\ldots&\beta_{m+k-1}\\
\beta_{m+1}&\beta_{m+2}&\ldots&\beta_{m+k}\\
\vdots&\vdots&&\vdots\\
\beta_{m+k-1}&\beta_{m+k}&\ldots&\beta_{m+2k-2}
\end{array}\right|
\label{eq:dhank3}
\eea
whose elements are given by 
\bea
\beta_{k}&=&-\alpha_{k}(t), \qquad k=1,2,\ldots , \nonumber\\
&=&\alpha_{k}(t), \qquad    k=0,-1,-2,\ldots ,
\label{eq:hank5}
\eea
then
the elements of the T-fraction are given by 
\beq
f_{n}=-\frac{v_{n-2}u_{n}}{v_{n-1}u_{n-1}}, \quad g_{n}=-\frac{v_{n-1}u_{n}}{v_{n}u_{n-1}}, \quad  n=2,3,4,\ldots . 
\label{eq:del1}
\eeq

From the expression for the $f_{n},g_{n}$ in terms of the coordinates and momenta given by (\ref{eq:dcont3}) and
the equation of motion (\ref{eq:dtl1}), it follows that
\beq
\frac{f_{1}(t+h)}{f_{1}(t)}=-g_{1}(t+h),\qquad \frac{g_{1}(t+h)}{g_{1}(t)}=\frac{1+\frac{f_{2}(t)}{g_{1}(t)}}{1+f_{1}(t)}
\label{eq:del2}
\eeq
and more generally
\beq
\fl 
\frac{f_{n+1}(t+h)}{f_{n+1}(t)}=\frac{g_{n+1}(t+h)}{g_{n}(t)},
\qquad  
\frac{g_{n+1}(t+h)}{g_{n+1}(t)}  
=\frac{1+\frac{f_{n+2}(t)}{g_{n+1}(t)}}{1+\frac{f_{n+1}(t)}{g_{n}(t)}},\quad  n\in\mathbb{N} .
\label{eq:del3}
\eeq

\begin{Pro}  
The two types of Hankel determinant defined by (\ref{eq:dhank2}) are related by    
\beq
v_{n}(t+h)=u_{n}(t), \qquad n=0,1,2,\ldots .
\label{eq:del4}
\eeq 
\end{Pro}  
\begin{prf} 
Substituting the expressions for   $f_{n},g_{n}$ given by (\ref{eq:del1}) into the first of (\ref{eq:del3}) one finds that
\beq
\frac{v_{n+1}(t+h) u_{n}(t)}{v_{n}(t+h)u_{n+1}(t)}=\frac{v_{1}(t+h)u_{0}(t)}{v_{0}(t+h)u_{1}(t)}.
\label{eq:del5}
\eeq 
But 
$u_{0}=H_{0}^{(2)}=1=H_{0}^{(1)}=v_{0}$ 
and
$u_{1}(t)=\beta_{1}=-h^{2}\exp(-x_{0}(t))$, 
$v_{1}(t)=\beta_{0}=-h^{2}\exp(-x_{0}(t-h))$.
Therefore the right hand side of (\ref{eq:del5}) equals unity and the result follows.$\,\,\Box$
\end{prf} 
\begin{Pro} 
When $ \exp(-x_{0}(t))=-\frac{t}{4} $, the Hankel determinants $u_n$ satisfy the bilinear relation 
\beq
\left(1-\frac{h^{2}t}{4}\right)\, u_{n}(t+h)u_{n}(t-h)=u_{n}(t)^{2}+u_{n+1}(t)u_{n-1}(t) 
\label{eq:ueq1}
\eeq 
for $n=1,2,\ldots$. 
\label{bilhank} 
\end{Pro} 
\begin{prf} 
Substituting for $g_{n}$ and $f_{n}$ from (\ref{eq:del1}) into the second of (\ref{eq:del3}) and then 
eliminating the dependence on the $v_{n}$
using (\ref{eq:del4}), the result follows from the fact that $u_{1}(t)=v_{1}(t+h)=\frac{h^2t}{4}$ and 
$u_{2}(t)
=-\frac{h^6}{64}(t^3-h^2t+4)$. 
$\,\,\Box$
\end{prf} 

\noindent {\bf Remark.} In \cite{KMNOY}, Hankel determinant solutions are 
given for a different (but gauge equivalent) bilinear form of the 
discrete 
Toda lattice equation, namely 
the equation 
$ 
\rho^{l+1}_n\rho^{l-1}_n-(\rho^{l}_n)^2=\varepsilon^2 
\rho^{l+1}_{n+1}\rho^{l-1}_{n-1}
$. 
Upon setting $\varepsilon^2=1$ and $t=(l-n)h$, 
$u_n(t)=\rho^{n+ht}_n/\rho^{ht}_0$ satisfies (\ref{eq:ueq1}) for 
a suitable choice of the initial condition $\rho^{ht}_0$ for $n=0$.     

It follows from the recurrence relations (\ref{eq:dhank1}) that when $\exp(-x_{0}(t))=-\frac{t}{4}$ the $\alpha_{j}(t)$, 
and hence also $u_{n}(t)$ and $v_{n}(t)$, 
are polynomials in $h,t$. We now renormalise the $u_{n}$ so that they are $\mathcal{O}(h^{0})$
as $h\rightarrow 0$.
\begin{Def} 
The discrete Yablonskii-Vorob'ev polynomials $Y_n(t,h)$ 
are defined by 
the bilinear recurrence  
\beq \fl 
h^{2}Y_{n+1}(t,h)Y_{n-1}(t,h)   
= (h^2 t-4)\, Y_{n}(t+h,h)Y_{n}(t-h,h) + 4Y_{n}(t,h)^2
\label{eq:dyab1}
\eeq
for $n\geq 1$, with $Y_0(t,h)=1$, $Y_1(t,h)=t$ as initial data.  
\end{Def}
\begin{Thm}
The discrete Yablonskii-Vorob'ev  
polynomials $Y_n(t,h)\in\mathbb{Z}[t,h^2]$ 
are given in terms of the Hankel determinants $u_n$ in (\ref{eq:dhank2}) by 
$Y_{n}(t,h) =(-1)^{n}(-\frac{h^{2}}{4})^{-n(n+1)/2}u_{n}(t)$ for $n=0,1,2,\ldots$. 
Each $Y_n$ is a  
monic polynomial of degree $n(n+1)/2$ in $t$, satisfying  
$Y_{n}(t,h)= y_{n}(t)+\mathcal{O}(h^{2})$ as $h\rightarrow 0$, 
where the $y_n(t)$ are the usual Yablonskii-Vorob'ev polynomials defined by (\ref{biltoda}) 
with $y_0=1$, $y_1=t$. 
\label{discyvdefn} 
\end{Thm}
\begin{prf} 
With $\alpha_1(t)=-h^2t/4$ it is easy to prove by induction from the first 
recursion in (\ref{eq:dhank1}) that $\alpha_j(t)=h^2P_j/4^j$ for $j\geq 1$, where $P_j\in\mathbb{Z}[t,h]$. 
Similarly, with $\alpha_0(t)=h^2(t-h)/4$ the second recursion in (\ref{eq:dhank1}) implies 
that $\alpha_j(t)=h^2\hat{P}_j/4^{-j+1}$ for $j\leq 0$, where $\hat{P}_j\in\mathbb{Z}[t,h]$. It follows 
from their definition in terms of the matrix elements (\ref{eq:hank5}) that 
the Hankel determinants $u_n$ are polynomials in $\mathbb{Q}[t,h]$, with powers of 4 as the only 
possible denominators of the coefficients. 
If we let $Y_{n}(t,h) =(-1)^{n}(-\frac{h^{2}}{4})^{-n(n+1)/2}u_{n}(t)$ for $n\in\mathbb{N}$ then we 
find $Y_0(t,h)=1$, $Y_1(t,h)=t$, and then 
the relation (\ref{eq:dyab1}) follows from Proposition \ref{bilhank}, by substituting for $u_{n}(t)$ in (\ref{eq:ueq1}) in terms of $Y_{n}(t,h)$. 
Since the $Y_n$ are uniquely defined by the recurrence  (\ref{eq:dyab1}) together with the given initial data, 
the formula in terms of renormalised Hankel determinants guarantees that they are polynomials in $t$, and the recurrence 
also implies that they are monic and of degree $d_n:=n(n+1)/2)$ in $t$.  However, further 
analysis is required to verify that there are no powers of $h$ or powers of $4$ in the denominator for this 
choice of normalisation. 

By induction, suppose that $Y_n(t,h)\in\mathbb{Z}[t,h]$, and hence is regular as $h\to 0$,
for $n=0,1,\ldots,N$ (which clearly holds for $N=1$).
Taking a Taylor expansion in $t$, with derivatives denoted by 
$Y_{N,(j)t}=\frac{\partial^j Y_N}{\partial t^j}$, we see that 
\[ 
\fl
\begin{array}{rcl} 
Y_{N}(t+h,h)Y_{N}(t-h,h)&=&\left(\sum_{k=0}^{[d_N/2]}\frac{h^{2k}Y_{N,(2k)t}(t,h)}{(2k)!}\right)^2 
-h^2\left(\sum_{k=0}^{[(d_N-1)/2]}\frac{h^{2k}Y_{N,(2k+1)t}(t,h)}{(2k+1)!}\right)^2 \\ 
&=& Y_N(t,h)^2 + h^2 \tilde{P}(t,h), 
\end{array} 
\] 
where $\tilde{P}(t,h)$ is a polynomial in $t$ and $h$ by the inductive hypothesis. Moreover, 
at leading order we have 
\[ 
\tilde{P}(t,h)=Y_N(t,0)Y_{N,tt}(t,0)-Y_{N,t}(t,0)^2 + \mathcal{O}(h).
\]   
Substituting this into the right hand side of (\ref{eq:dyab1}) and dividing by $h^2$ we have 
\beq \label{taylor} 
Y_{N+1}(t,h)Y_{N-1}(t,h) = t Y_{N}(t+h,h)Y_{N}(t-h,h)-4 \tilde{P}(t,h), 
\eeq 
from which it follows that $Y_{N+1}(t,h)$ is regular as $h\to 0$, and hence lies in 
$\mathbb{Q}[t,h]$ with at worst powers of 4 as denominators of its coefficients. 
Now for some $K\geq 0$ 
we can write $Y_{N+1}(t,h)=\hat{Y}_{N+1}(t,h)/4^K$ 
where $\hat{Y}_{N+1}\in\mathbb{Z}[t,h]$ 
with $4\not |\, \hat{Y}_{N+1}$. 
If we multiply both sides of (\ref{eq:dyab1}) through by 
$4^K$ we see that if $K>0$ then we have $4| h^2 \hat{Y}_{N+1}Y_{N-1}$; but then since 
$Y_{N-1}$ is monic in $t$ it is clear that 
$2\not |\, Y_{N-1}(t,h)$ we must have $4|\hat{Y}_{N+1}$, 
which is a contradiction. Hence $K=0$ and $Y_{N+1}\in\mathbb{Z}[t,h]$ as required.    

Setting $h\to -h$ in the recurrence (\ref{eq:dyab1}) it also follows immediately 
by induction that $Y_n(t,h)=Y_n(t,-h)$ for all $n$, so in fact we have polynomials in 
$\mathbb{Z}[t,h^2]$. Thus we can write $Y_{n}(t,h)=y_n(t)+\mathcal{O}(h^2)$ where 
$y_0(t)=Y_0(t,h)=1$ and $y_1(t)=Y_1(t,h)=t$, and using the leading order part of  
$\tilde{P}(t,h)$ in (\ref{taylor}) we find that for $n\geq 1$ the $y_n$ 
satisfy 
\[ 
y_{n+1}y_{n-1}=ty_n^2-4y_n\ddot{y}_n+4\dot{y}_n^2, 
\] 
which is precisely the defining relation (\ref{biltoda}) for the 
usual Yablonskii-Vorob'ev polynomials.$\,\Box$
\end{prf} 

The first few discrete Yablonskii-Vorob'ev polynomials are
\bea
Y_0(t,h)&=& 1, \nonumber   \\
Y_{1}(t,h)&=& t, \nonumber   \\
Y_{2}(t,h)&=&t^3+4-h^2 t, \nonumber\\
Y_{3}(t,h)&=&t^6+20t^3-80+h^2(4t-5t^4)+4t^2h^4, \nonumber\\
Y_{4}(t,h)&=&t^{10}+60t^{7}+11200t-h^2(15t^8+252t^5+3360t^2)+\nonumber \\&&+h^4(63t^6+480t^3+576)-h^6(85t^4+288t)+36h^8t^2.
\label{eq:dyab3}
\eea
These examples clearly reduce to the usual Yablonskii-Vorob'ev polynomials 
(\ref{pols}) when $h=0$.

\section{Zeroes of discrete Yablonskii-Vorob'ev polynomials}

In the continuum case it has been shown that the zeroes of $y_{n}(t)$ are simple and are not zeroes of 
$y_{n+1}(t)$ \cite{ok, tan}. Here we prove an analogous result in the discrete case, by 
considering the roots of $Y_n(t,h)$ as a polynomial in $t$, that is $t_0=t_0(h)$ such that 
$Y_n(t_0(h),h)=0$. To begin with we require a simple observation. 
\begin{Lem}\label{eq:zer1}
The discrete Yablonskii-Vorob'ev polynomials never vanish at $t=4/h^2$. More 
precisely,    
\[
\fl 
\qquad Y_{n}\left(\frac{4}{h^2},h\right)=\left(\frac{4}{h^2}\right)^{n(n+1)/2} \neq 0 
\qquad for \, all \, n\in\mathbb{N}.  
\]
\end{Lem}
\begin{prf}
The result is trivially true when $n=0,1$.
Assume that it holds for $n=1,2,\dots,N$. Then from (\ref{eq:dyab1})
$h^2Y_{N+1}(4/h^2,h)Y_{N-1}(4/h^2,h)=4Y_{N}(4/h^2,h)^2\neq 0$, 
and the result follows 
by induction. 
$\,\,\Box$  
\end{prf}

From Theorem \ref{discyvdefn} we know that $Y_n(t+h,h)-Y_n(t,h)=h\dot{y}_n(t) +\mathcal{O}(h^2)$, and the 
zeroes of $Y_n(t,h)$ in $t$ differ from those of $y_n(t)$ by $\mathcal{O}(h^2)$, so (because 
the zeroes of $y_n$ are simple) we expect that $Y_n(t+h,h)$ should not vanish where $Y_n(t,h)$ does. Indeed 
this turns out to be the case. In the next section we shall see that 
the polynomials $Y_N$ are tau-functions for a discrete PII equation, and hence 
satisfy many other bilinear identities in addition to (\ref{eq:dyab1}). To prove 
the following result we make use of one such identity, namely (\ref{eq:qc}) below.   
\begin{Thm} \label{zerothm}
For any $n=1,2,\ldots$,  
if $t_0$ is a zero of $Y_n$,
so that $Y_{n}(t_{0},h)=0$, then 
$Y_{n}(t_{0}\pm h,h)\neq0$ and $Y_{n+1}(t_{0},h)\neq 0$. 
\end{Thm}
\begin{prf}
It is true for $n=1$ by inspection. Suppose  the result is true for $n=1,2,\ldots,N-1$. 
Now assume that both $Y_{N}(t_{0},h)=0$ and $Y_{N}(t_{0}+h,h)=0$. 
Using the identity (\ref{eq:qc}) with $\ta_{n}=Y_{n}(t,h)$,
and setting $n=N-1$, $t=t_{0}$, we have 
\[ 
\fl  
\begin{array}{lcl}  
Y_{N}(t_{0}+h,h)Y_{N-2}(t_{0},h) & - & Y_{N}(t_{0},h)Y_{N-2}(t_{0}+h,h) 
\\
& = & (2N-1)hY_{N-1}(t_{0}+h,h)Y_{N-1}(t_{0},h).
\end{array} 
\] 
By the assumption, the left hand side vanishes, and $2N-1\neq 0$ for 
$N\in\mathbb{Z}$ which means that $Y_{N-1}(t_{0},h)$ or $Y_{N-1}(t_{0}+h,h)=0$; 
but this contradicts the inductive hypothesis, 
so $Y_{N}(t_0+h,h)\neq 0$ whenever $Y_{N}(t_{0},h)=0$ (and similarly for 
$Y_{N}(t_0-h,h)$). Now if $Y_{N}(t_{0},h)=0$, 
then from (\ref{eq:dyab1}) we see that 
\[  
h^2Y_{N-1}(t_{0},h)Y_{N+1}(t_0,h) = 
(h^2t_{0}-4)Y_{N}(t_{0}+h,h)Y_{N}(t_{0}-h,h)
\]
If we now assume that also $Y_{N+1}(t_0,h)=0$, then the left hand side of the 
above vanishes, and since $h^2t_{0}-4\neq 0$ by Lemma \ref{eq:zer1} then 
$Y_{N}(t_{0}+h,h)=0$ or $Y_{N}(t_{0}-h,h)=0$, which is a contradiction.  
Therefore the result holds for 
$n=N$, hence for all $n$ by induction.
$\,\,\Box$
\end{prf}

Clarkson and Mansfield have numerically studied 
the location of zeroes of $y_{n}(t)$ for low values
of $n$ \cite{pet2}. They found that for a given $n$ they occupy approximate triangular arrays in the complex plane and 
that the zeroes of $y_{n}(t)$ \emph{interlace} in a certain sense with those of $y_{n+1}(t)$.
Since the zeroes of $Y_n(t,h)$ are the same as those of $y_n(t)$ 
up to $\mathcal{O}(h^2)$, the same picture holds for sufficiently small $h$, and   
our numerical observations show that the same qualitative behaviour persists  
for values of $h$ up to at least order of unity.

\section{Alternate discrete Painlev\'{e} II}

The second Painlev\'{e} equation ($P_{II}$) is 
\beq
\ddot{q}=2q^3+tq+\alpha,
\label{eq:p2a}
\eeq
where $\alpha$ is a constant. This second order differential equation is 
equivalent to a first order system introduced by Okamoto \cite{okamoto}, namely 
\beq
\begin{array}{rcl} 
\dot{q}&=&p -q^2 -\frac{t}{2} \\
\dot{p}&=&2qp+\ell, \qquad \ell = \alpha + \frac{1}{2}.
\end{array} 
\label{p2sys}
\eeq
Rational solutions of (\ref{eq:p2a}) arise when 
$\alpha=n \in \mathbb{Z}$. However, in this and subsequent sections $n$ need not be 
an integer unless stated explicitly otherwise, and we use the parameters 
$n$ and $\alpha$ interchangeably. We also find it convenient to 
use the shifted parameter $\ell = n+1/2$ as above, which 
fixes a point in the $A_1$ root space. This is the parameter used by 
Okamoto to label solutions of $P_{II}$ and the corresponding tau-functions, 
but we stick with the label $n$ to maintain contact with the preceding results.
$P_{II}$ may be put into bilinear form by setting 
\beq
q_{n}=\frac{d}{dt}\log\left(\frac{\ta_{n-1}(t)}{\ta_{n}(t)}\right).
\label{eq:bla}
\eeq 
Then(\ref{eq:p2a}) with $\alpha=n$ is equivalent to the bilinear equations 
\beq
\begin{array}{rcl} 
    D_{t}^{2}\ta_{n}\cdot  \ta_{n-1}&=&F(t)\, \ta_{n} \ta_{n-1}, \\
(D_{t}^{3}-tD_{t}+n)\ta_{n}\cdot \ta_{n-1}&=&3F(t)\,D_t \ta_{n}\cdot  \ta_{n-1} ,
\end{array} 
\label{eq:blb}
\eeq
where $D_{t}$ is the usual Hirota operator, and the function $F(t)$ is arbitrary. 
By a gauge transformation on the tau-functions, that is $\tau_n \to G(t) \tau_n$ for all $n$, 
the function $F$ can be set to zero in (\ref{eq:blb}) without loss of generality. 
 
The B\"{a}cklund transformation linking a solution of (\ref{eq:p2a}) 
for $\alpha=n$ with that for $\alpha=n+1$ is  
\beq 
  q_{n+1}=-q_{n}-\frac{\ell}{p_n}, \qquad p_n=\dot{q}_{n}+q_{n}^2+t/2,
\label{eq:bac1}
\eeq
and the two solutions are related by 
\beq
\dot{q}_{n}+q_{n}^2=-\dot{q}_{n+1}+q_{n+1}^2 
\label{eq:bac2}
\eeq
(which can be seen as being inherited from the Miura transformation for KdV). 
From (\ref{eq:blb}) and (\ref{eq:bac1}) 
it may be proved that, with the same choice of gauge that fixes $F\equiv 0$ in (\ref{eq:blb}), 
any three adjacent tau-functions 
$\ta_n$, $\ta_{n\pm 1}$ satisfy the bilinear relations  
\beq
D_{t}\ta_{n+1}\cdot\ta_{n-1}=2\ell\,\ta_{n}^2 
\label{eq:blc}
\eeq
and
\beq
\ta_{n+1}\ta_{n-1}=t\ta_{n}^2-2D_t^2 \ta_{n}\cdot\ta_n. 
\label{eq:bld}
\eeq
With this choice of gauge, the equation (\ref{eq:bld}) is the bilinear form 
of the Toda lattice. As we have already mentioned (cf. equation (\ref{biltoda}) in the Introduction), 
the latter relation is the defining recurrence for the Yablonksii-Vorob'ev polynomials; 
with $\tau_n=y_n$ they satisfy the other bilinear equations (\ref{eq:blc}) and (\ref{eq:blb}) for 
$F=0$, and hence determine the rational solutions of $P_{II}$.

Several discretisations of $P_{II}$ have been studied 
in recent years including a discrete bilinear form \cite{sats}
which tends to (\ref{eq:blb}) in the continuum limit. The discrete 
bilinear form allowed the construction  
of rational solutions to this discrete $P_{II}$, given 
in terms of ratios of determinants of Jacobi-Trudi type \cite{kko}, while  
solutions in terms of discrete Airy functions were found in \cite{kaj,kaj2}. 
A hierarchy of higher order analogues of this discrete $P_{II}$ ($dP_{II}$) have been given 
by Cresswell and Joshi, based on a Lax pair. 
However, due to the non-uniqueness of the discretisation process, 
there are many 
other  $dPII$ equations with analogous properties. 
For example,  
four different $dP_{II}$ equations are mentioned in 
\cite{coal}, three of which are $q$-type ($q$-$P_{II}$). Discrete Airy function solutions of 
one such  $q$-$P_{II}$ equation are found in \cite{tamizhmani}, while another one is considered 
with its ultra-discretisation in \cite{isojima}, along with yet one more distinct 
$q$-$P_{II}$. In \cite{HKW} a detailed study of a 
$q$-$P_{II}$ in Sakai's classification \cite{sakai} 
has been performed. Another alternative discretisation, which is 
relevant here, is the so called 
{\it alternate} $dP_{II}$ ($alt$-$dP_{II}$) which was studied in 
\cite{FSKGR}, and appeared in connection with the original 
$dP_{II}$ in \cite{AGT}.

Here we consider 
the $alt$-$dP_{II}$ equation in the 
form of the second order difference equation 
\beq 
\frac{\frac{h^3}{2}m -2}{\ovl{g} g +1} 
+  
\frac{\frac{h^3}{2}(m-1) -2}{g\unl{g}  +1}  
= 
-g+\frac{1}{g} + \frac{h^3}{2}(m-\ell )-2 
\label{eq:dp2a}
\eeq 
with the notation $g=g_n(t)$, $\ovl{g}_{n}=g_{n}(t+h,h)$, 
$\unl{g}_{n}=g_{n}(t-h,h)$ 
\footnote{The introduction of $m$ in place of $t$ is not essential, but it suggests that $m$ and $\ell$ 
might be considered on the same footing, so that $w_{\ell ,m}\equiv g_n(t)$ 
should provide particular solutions of a suitable partial difference equation. 
See \cite{FSKGR} for a connection with the lattice modified BSQ equation.}.   
This is referred to as 
a discrete $P_{II}$ equation due to the fact that (\ref{eq:p2a}) 
arises from it by setting $g(t,h)=-1+hq(t,h)$ and taking the continuum limit $h\to 0$ 
(with parameter $\alpha = \ell -1/2$ as usual). If we set $g_n(t,h)=x_m$ 
with $z_m = h^3m/2-2$ and $\mu = -h^3\ell /2$ then this is 
equation (1.3) in \cite{FSKGR} (except that we have $m$ in place of $n$).

The $alt$-$dP_{II}$ equation (\ref{eq:dp2a}) 
has  rational solutions for integer values of the parameter $n=\ell -1/2$, provided by suitable ratios 
of the discrete Yablonskii-Vorob'ev 
polynomials $Y_{n}(t,h)$. In order to prove this, we must derive appropriate discrete analogues 
of the bilinear identities (\ref{eq:blb}), (\ref{eq:blc}) and (\ref{eq:bld}). We start by presenting 
(\ref{eq:dp2a}) in the form of a system, which makes it more manageable. 

\begin{Lem} 
The $alt$-$dP_{II}$ equation 
(\ref{eq:dp2a}) for $\ell = n+\frac{1}{2}$, $m=t/h$  
is equivalent to the first order system 
\beq
\begin{array}{rcl} 
\ovl{g}_n & = & (-2+h^2p_n)/(4+2g_n-h^2(g_np_n+t)),  
\\ 
\ovl{p}_n & = & (p_n -\ell h\ovl{g}_n)/\ovl{g}_n^2. 
 \label{dp2sys} 
\end{array}
\eeq 
\end{Lem}
\begin{prf} 
By rearranging the first equation in (\ref{dp2sys}) we have 
\beq \label{pdef} 
p_n=\frac{1}{h^2}\left( \frac{(4-h^2t)\ovl{g}_n}{1+g_n\ovl{g}_n} +2 \right),  
\eeq 
and by shifting $t\to t+h$ the latter provides a formula for $\ovl{p}_n$ 
in terms of $\ovl{g}_n$ and $\ovl{\ovl{g}}_n$, and then substituting this 
into the left hand side of the second equation (\ref{dp2sys}), with $p_n$ given 
by (\ref{pdef}) on the right hand side, gives a relation between 
$g_n$, $\ovl{g}_n$ and $\ovl{\ovl{g}}_n$. After rearranging, shifting $t\to t-h$, 
and setting $t=mh$, this relation is precisely (\ref{eq:dp2a}). 
Conversely, given the $alt$-$dPII$ equation 
(\ref{eq:dp2a}), we can define $p_n$ by the formula (\ref{pdef}), 
which is equivalent to the first equation in (\ref{dp2sys});  the 
second equation for $\ovl{p}_n$ then follows immediately from $alt$-$dPII$.  
$\,\,\Box$ 
\end{prf} 

The B\"{a}cklund transformation  
for the $alt$-$dP_{II}$ equation  (\ref{eq:dp2a}) is most easily derived 
starting from the analogue of (\ref{eq:bac2}), and the following results 
are easily verified by direct calculation. 

\begin{Lem} \label{dmiura}
If $g_{n+1}$ is given by 
\beq \label{gbt} 
g_{n+1}=\frac{p_n}{\ell h +g_n p_n}, 
\eeq 
in terms of $g_n$ and $p_n$ satisfying
(\ref{dp2sys}), 
then the identity
\beq
\frac{1}{\ovl{g}_{n}}+g_{n}=\frac{1}{g_{n+1}}+\ovl{g}_{n+1}
\label{eq:dp2d}
\eeq
holds. 
\end{Lem}
\begin{Cor} \label{miuracor}
The quantity $g_{n+1}$, defined by (\ref{gbt}) with  $p_n$ given by 
(\ref{pdef}), satisfies (\ref{eq:dp2a}) with $\ell\to \ell +1$. Equivalently, 
the equation (\ref{gbt}) and the relation 
\beq \label{pbt} 
p_{n+1}= \frac{1}{h^2}\left( \frac{(4-h^2t)\ovl{g}_{n+1}}{1+g_{n+1}\ovl{g}_{n+1}} +2 \right)
\eeq 
together constitute a B\"{a}cklund transformation  
for the $alt$-$dP_{II}$ system  (\ref{dp2sys}). 
\end{Cor} 
\begin{Cor} 
The  B\"{a}cklund transformation for the 
$alt$-$dP_{II}$ system, given by the formulae (\ref{gbt}) and (\ref{pbt}), has  
the following consequences:
\beq \label{gform} 
g_{n+1}\unl{p}_n=g_np_n; 
\eeq 
\beq \label{qcform} 
\frac{1}{\ovl{g}_n}-\ovl{g}_{n+1}=\frac{\ell h}{p_n}; 
\eeq 
\beq \label{qnform} 
\ovl{g}_{n+1}g_n-\frac{1}{g_{n+1}\ovl{g}_n}=
-\frac{\ell h}{p_n}\left(g_n+\frac{1}{\ovl{g}_n}\right). 
\eeq  
\end{Cor} 
\noindent {\bf Remark.} With $g_n=-1+hq_n$, the identity (\ref{eq:bac2}) arises 
as the continuum limit of (\ref{eq:dp2d}), and the formula (\ref{eq:bac1}) 
arises from (\ref{gbt}), as $h\to 0$. Similarly, the system (\ref{p2sys}) 
is the continuum limit of (\ref{dp2sys}). 
Nijhoff {\it et al.} derived equivalent formulae of Miura/Schlesinger type  
for the B\"{a}cklund transformation of $alt$-$dP_{II}$   
by making use of a variable $y_n$ (see equation (5.1) in \cite{FSKGR}), 
which (modulo rescaling and replacing $n$ by $m$) 
is analogous to $p_n$ defined by (\ref{pdef}).



We now describe the tau-functions 
for the $alt$-$dP_{II}$ equation, which satisfy analogues of  (\ref{eq:blb}).  

\begin{Pro} \label{taufneqns}
Up to a choice of gauge, every solution of (\ref{eq:dp2a}) is 
specified by a pair of tau-functions $\ta_n(t,h)$, $\ta_{n-1}(t,h)$ 
via the formula
\beq
g_{n}(t,h)=-\frac{\ta_{n-1}(t-h,h)\ta_{n}(t,h)}{\ta_{n-1}(t,h)\ta_{n}(t-h,h)}, 
\label{eq:qa}
\eeq
where the tau-functions 
satisfy the bilinear equation 
\beq
\ovl{\ta}_{n}\unl{\ta}_{n-1}+\unl{\ta}_{n}\ovl{\ta}_{n-1}=2\ta_{n}\ta_{n-1} 
\label{eq:qb}
\eeq 
and the quadrilinear (degree four) equation 
\beq 
\fl 
\begin{array}{r} 
(4-mh^3)\unl{\ta}_{n-1}\unl{\ta}_{n}(\ovl{\ta}_{n-1}\unl{\ta}_{n}-\unl{\ta}_{n-1}\ovl{\ta}_{n}) 
+ (4-(m-1)h^3){\ta}_{n-1}{\ta}_{n}({\ta}_{n-1}\unl{\unl{\ta}}_{n}-\unl{\unl{\ta}}_{n-1}{\ta}_{n})
 \\
+8(\unl{\ta}_{n-1}^2\ta_n^2 
-\ta_{n-1}^2\unl{\ta}_{n}^2) 
-4n h^3 \unl{\ta}_{n-1}\unl{\ta}_{n}{\ta}_{n-1}{\ta}_{n}=0 , 
\end{array} 
\label{eq:qbquad}
\eeq
with $m=t/h$, $n=\alpha =\ell-1/2$ and 
$\ta_{n}=\ta_{n}(t,h),\ovl{\ta}_{n}=\ta_{n}(t+h,h),\unl{\ta}_{n}=\ta_{n}(t-h,h)$, etc. 
\end{Pro} 
\begin{prf}
Upon substituting the tau-function expression (\ref{eq:qa}) into 
(\ref{eq:dp2a}) and clearing denominators, a relation of degree eight results, 
which can be simplified somewhat by rewriting it in terms of the symmetric/antisymmetric 
quadratic quantities $A_{\pm}=\ovl{\ta}_{n-1}\unl{\ta}_n \pm \unl{\ta}_{n-1}\ovl{\ta}_n$ 
and $\unl{A}_{\pm}={\ta}_{n-1}\unl{\unl{\ta}}_n \pm \unl{\unl{\ta}}_{n-1}{\ta}_n$. 
In general, for any choice of tau-functions the bilinear  equation 
\[ 
\ovl{\ta}_{n}\unl{\ta}_{n-1}+\unl{\ta}_{n}\ovl{\ta}_{n-1}=2\hat{F}\,\ta_{n}\ta_{n-1} 
\] 
holds, for some function $\hat{F}=\hat{F}(t,h)$, but by applying a gauge 
transformation $\ta_n\to \hat{G}\, \ta_n$, $\ta_{n-1}\to \hat{G}\, \ta_{n-1}$ with 
$\ovl{\hat{G}}\unl{\hat{G}}/\hat{G}^2=\hat{F}$ the function $\hat{F}$ can be removed to yield the bilinear 
equation (\ref{eq:qb}). With this choice of gauge, the remaining terms in the degree eight 
relation factorise to yield the quadrilinear equation (\ref{eq:qbquad}), 
and conversely if these two tau-function equations hold then 
$g_n$ given by (\ref{eq:qa}) is a solution of (\ref{eq:dp2a}) 
for $\ell =n+1/2=\alpha +1/2$. 
$\,\,\Box$ 
\end{prf} 

\noindent {\bf Remark.} 
The existence of a quadrilinear relation between a pair of tau-functions 
is mentioned in section 4 of \cite{FSKGR}, where a third tau-function is 
introduced to obtain purely bilinear relations (cf. Theorem  
\ref{normalise} below).  


It is easy to see that (\ref{eq:qb}) tends to the first of (\ref{eq:blb}) in the continuum
limit (with the gauge chosen so that $F=0$). Although the second relation (\ref{eq:qbquad}) between the 
two tau-functions is of overall degree four, it still produces the second bilinear differential equation 
(\ref{eq:blb}) in the continuum limit (provided that the first one also holds). 
In order to work with purely bilinear equations in the discrete case, we must 
consider three adjacent tau-functions $\ta_n$, $\ta_{n\pm 1}$.  


\begin{Thm}\label{normalise}  
Up to a choice of gauge, every solution of the  $alt$-$dP_{II}$ system  (\ref{dp2sys}) is specified 
by three tau-functions $\ta_{n-1}(t)$, $\ta_n(t)$, $\ta_{n+1}(t)$, 
with $g_n$ given by (\ref{eq:qa}) and 
\beq \label{ptau} 
p_{n}=\frac{\ta_{n-1}\ta_{n+1}}{2\ta_{n}^2},  
\eeq 
where the tau-functions satisfy (\ref{eq:qb}) as well as 
\beq
\ovl{\ta}_{n+1}\unl{\ta}_{n}+\unl{\ta}_{n+1}\ovl{\ta}_{n}=2\ta_{n+1}\ta_{n} 
\label{eq:qbbt}
\eeq
and 
\beq 
h^2\ta_{n+1}\ta_{n-1}=(h^2t-4)\ovl{\ta}_n\unl{\ta}_n+4\ta_n^2. 
\label{eq:qd}
\eeq
With this choice of normalisation the identities  
\beq
 \ovl{\ta}_{n+1}\ta_{n-1}-\ta_{n+1}\ovl{\ta}_{n-1} = 2\ell h\ovl{\ta}_{n}\ta_{n}
\label{eq:qc}
\eeq
and
\beq
\ovl{\ta}_{n+1}\unl{\ta}_{n-1}-\unl{\ta}_{n+1}\ovl{\ta}_{n-1}=4\ell h\ta_{n}^2
\label{eq:qn}
\eeq
also hold. For this choice of gauge, these purely bilinear relations are compatible with the B\"acklund 
transformation (\ref{gbt}) 
for $alt$-$dP_{II}$, in the sense that  $\ta_n$ and $\ta_{n+1}$ satisfy (\ref{eq:qbquad}) for $n\to n+1$, and 
\[
g_{n+1}(t,h)=-\frac{\ta_{n}(t-h,h)\ta_{n+1}(t,h)}{\ta_{n}(t,h)\ta_{n+1}(t-h,h)}
\]
satisfies (\ref{eq:dp2a}) with $\ell\to\ell +1$. 
\end{Thm} 
\begin{prf}
If  a solution $g_n,p_n$ of (\ref{dp2sys}) is given by the expressions (\ref{eq:qa}) 
and (\ref{ptau}) respectively, and the gauge is fixed by  (\ref{eq:qb}), then the 
latter implies that 
\[ 
\fl 
g_n + \frac{1}{\ovl{g}_n}  =  -2\frac{\ta_n^2}{\ovl{\ta}_n\unl{\ta}_n}  
                           =   \frac{1}{g_{n+1}}+\ovl{g}_{n+1} 
\]
by Lemma \ref{dmiura}, where $g_{n+1}$ given by (\ref{gbt}) is a solution 
of (\ref{eq:dp2a}) with $\ell\to\ell +1$. The first equality 
above implies (\ref{eq:qd}), while the relation (\ref{gform}) implies 
that $g_{n+1}$ is given in terms of tau-functions by the formula  
(\ref{eq:qa}) with $n\to n+1$, and hence (\ref{eq:qbbt}) follows from 
the second equality above. The bilinear identities (\ref{eq:qc}) and 
(\ref{eq:qn}) then hold as a consequence of the relations 
(\ref{qcform}) and (\ref{qnform}) respectively.  By Proposition  
\ref{taufneqns}, the given choice of gauge implies that 
the pair $\ta_{n-1}$, $\ta_n$ satisfy (\ref{eq:qbquad}), and the fact that  
$g_{n+1}$ is a solution of $alt$-$dPII$ with the parameter shifted implies that 
the pair $\ta_{n}$, $\ta_{n+1}$ also satisfy this quadrilinear equation 
with $n\to n+1$.       

Conversely, suppose that $g_n$ is defined in terms of tau-functions by (\ref{eq:qa}), $g_{n+1}$ 
is defined by the same relation for $n\to n+1$, and $p_n$ is defined by (\ref{ptau}), 
where the tau-functions satisfy the three relations (\ref{eq:qb}),  (\ref{eq:qbbt}) 
and (\ref{eq:qd}).  The identities (\ref{pdef}) and (\ref{eq:dp2d}) follow 
immediately. Furthermore, from (\ref{eq:qb}) and  (\ref{eq:qbbt}) it is 
clear that 
\[ 
\frac{\ta_{n-1}}{\ta_n}\left(\frac{\ovl{\ta}_{n+1}}{\ovl{\ta}_{n}} + \frac{\unl{\ta}_{n+1}}{\unl{\ta}_{n}}\right) 
=
\frac{\ta_{n+1}}{\ta_n}\left(\frac{\ovl{\ta}_{n-1}}{\ovl{\ta}_{n}} + \frac{\unl{\ta}_{n-1}}{\unl{\ta}_{n}}\right), 
\] 
which implies that 
\[ 
\frac{\ovl{\ta}_{n+1}\ta_{n-1}-\ta_{n+1}\ovl{\ta}_{n-1}}{\ovl{\ta}_{n}\ta_{n}} = 
\frac{{\ta}_{n+1}\unl{\ta}_{n-1}-\unl{\ta}_{n+1}{\ta}_{n-1}}{{\ta}_{n}\unl{\ta}_{n}}.
\] 
Therefore $(\ovl{\ta}_{n+1}\ta_{n-1}-\ta_{n+1}\ovl{\ta}_{n-1})/(2h\ovl{\ta}_{n}\ta_{n})$ is 
independent of $t$, and if we denote this by $\ell$, then we have the bilinear 
equation (\ref{eq:qc}), which implies that (\ref{qcform}) also holds. Solving 
(\ref{qcform}) for $\ovl{g}_{n+1}$ gives an expression in terms of 
$p_n$ and $\ovl{g}_n$, which in turn means that $\ovl{g}_{n+1}$ can be 
written in terms of $g_n$ and $\ovl{g}_n$ using (\ref{pdef}). By shifting 
$t\to t-h$, this gives a formula for $g_{n+1}$ in terms of $\unl{g}_n$ 
and $g_n$, and then substituting for $\ovl{g}_{n+1}$ and $g_{n+1}$ 
in (\ref{eq:dp2d}) yields the $alt$-$dP_{II}$ equation (\ref{eq:dp2a}). 
It then follows  that  $g_n,p_n$ satisfy the system (\ref{dp2sys}).  
$\,\,\Box$
\end{prf}  

\begin{Thm} \label{ratthm}
For parameter $\ell = n+\frac{1}{2}$ with $n\in\mathbb{Z}$ the $alt$-$dP_{II}$ equation (\ref{eq:dp2a}) 
has rational solutions given in terms of the discrete Yablonskii-Vorob'ev polynomials by 
\[ 
g_n = -\frac{Y_{n-1}(t-h,h)Y_n(t,h)}{Y_{n-1}(t,h) Y_{n}(t-h,h)}, 
\] 
where the polynomials are extended to negative indices $n$ by setting $Y_{-n}=Y_{n-1}$ for 
$n\in\mathbb{N}$. As well as the defining recurrence (\ref{eq:dyab1}), the 
relations  (\ref{eq:qb}), (\ref{eq:qbquad}), (\ref{eq:qc}) and (\ref{eq:qn}) 
are satisfied by $\ta_n(t,h)=Y_n(t,h)$ for all $n\in\mathbb{Z}$. 
\end{Thm} 
\begin{prf}
When $\ell = 1/2$, the equation (\ref{eq:dp2a}) has the trivial constant solution 
$g_0(t,h)=-1$, which can be obtained by setting $\ta_0=Y_0=1=Y_{-1}=\ta_{-1}$ in 
(\ref{eq:qa}), and from (\ref{pdef}) we have $p_0=t/2$ which 
gives $\ta_1 = Y_1=t$ by (\ref{ptau}). It is easy to verify that each of the 
bilinear equations (\ref{eq:qb}),  (\ref{eq:qbbt}) 
and (\ref{eq:qd}) is satisfied by these tau-functions. By applying the B\"acklund 
transformation (\ref{gbt}) repeatedly (both forwards and backwards) 
a doubly infinite sequence of rational solutions $\{\, g_n \,\}_{n\in\mathbb{Z}}$ 
is obtained. Then by Theorem \ref{normalise}, since the B\"acklund 
transformation is compatible with the choice of gauge, it follows by induction 
that the corresponding tau-functions satisfy the identities  
(\ref{eq:dyab1}), (\ref{eq:qb}), (\ref{eq:qbquad}), (\ref{eq:qc}) and (\ref{eq:qn}) 
for all $n\in\mathbb{Z}$. Since the Yablonskii-Vorob'ev polynomials are defined 
by (\ref{eq:qd}) with $\ta_0=1$, $\ta_1=t$, it follows that 
this particular sequence of tau-functions is given by $\ta_n(t,h) = Y_n(t,h)$ 
for all $n\in\mathbb{N}$. The fact that this relation can be 
consistently extended to negative $n$ follows from the observation that 
all of the tau-function identities in Proposition \ref{taufneqns}
and   Theorem \ref{normalise} are invariant under 
$n\to -n-1$, $\ell \to -\ell$. 
$\,\,\Box$
\end{prf} 

\noindent {\bf Remark.} The simplest rational solutions of $alt$-$dP_{II}$ 
(corresponding to $n=0,\pm 1$) 
are described in section 6 of the paper \cite{FSKGR} by 
Nijhoff {\it et al.}, where it is indicated how the above sequence of 
rational solutions can be generated recursively via the B\"acklund transformation, 
but no closed form for these rational solutions is given in that work.

The fact that the $alt$-$dP_{II}$ equation can be derived from 
a sequence of B\"acklund transformations applied to  
solutions of $P_{III}$ provides a relation between 
the discrete Yablonskii-Vorob'ev polynomials and 
the Umemura polynomials for $P_{III}$. The third Painlev\'e equation, 
$P_{III}$, is given by 
\beq \label{p3} 
w''=\frac{(w')^2}{w}-\frac{w'}{x}+\frac{1}{x}(\alpha w^2 +\beta ) 
+\gamma w^3 +\frac{\delta}{w}, 
\eeq 
with the prime $'$ denoting $d/dx$,
and without loss of generality (by rescaling the independent variable $x$) 
if $\gamma\delta\neq 0$ the latter two parameters can be fixed 
as $\gamma = -\delta =1$. 
(Note that the parameter $\alpha$ in (\ref{p3}) should not be 
confused with the parameter $\alpha$ in $P_{II}$.) 
B\"acklund transformations for $P_{III}$ 
(cf. section 2 in \cite{FSKGR}) can be used to relate three adjacent 
solutions $w_n=w(x;\alpha , \beta )$ and $w_{n\pm 1}=w(x;\alpha\pm 2 , \beta\pm 2 )$ 
and if we set $g_n = -1/w_n$ then this contiguity relation can be 
written in the form of the $alt$-$dP_{II}$ equation 
\beq \label{adp2s} 
\frac{z_n}{g_{n+1}g_n-1}+ 
\frac{z_{n-1}}{g_ng_{n-1}-1}+\frac{x}{2}\left( g_n +\frac{1}{g_n}\right) +z_n +\mu =0, 
\eeq  
where 
\beq \label{zmu} 
z_n=(\alpha + \beta +2)/4, \qquad \Delta z_n:= z_{n+1}-z_n=1, 
\qquad \mu = (\beta -  \alpha -2)/4. 
\eeq 
(To compare with equation (2.5) in \cite{FSKGR}, set $g_n=ix_n$, $x=t$ in the above.)
For a certain set of parameter values, 
$P_{III}$ has rational solutions which are described by the 
following result. 

\begin{Thm}\label{kajmas} (Kajiwara \& Masuda \cite{KM}) 
For parameters 
\beq\label{abe} 
\alpha = 2n+2\nu -1,  
\qquad \beta = 2n -2\nu +1, \qquad n\in\mathbb{Z} 
\eeq 
and $\gamma = -\delta = 1$, 
the third 
Painlev\'e equation (\ref{p3}) 
has rational solutions $w(x;\alpha , \beta )=w_n$ given by 
\beq\label{wform} 
w_n=\frac{\mathcal{D}_{n} (x,\nu -1 ) \mathcal{D}_{n-1}(x,\nu )}
{\mathcal{D}_n(x,\nu )\mathcal{D}_{n-1}(x,\nu -1 )} 
\eeq 
where the polynomial $\mathcal{D}_n$ is given by a determinant of Jacobi-Trudi type, 
$$ 
\mathcal{D}_n (x,\nu ) = \left| \begin{array}{cccc} 
p_n & p_{n+1} & \ldots & p_{2n-1} \\ 
 p_{n-2} & p_{n-1} & \ldots & p_{2n-3} \\ 
\vdots & \vdots & \ddots  & \vdots \\ 
p_{-n+2} & p_{-n+3} & \ldots & p_{1} 
\end{array}\right|
$$ 
with $p_k=p_k(x,\nu )$ defined by the generating function 
$$ 
\sum_{k=0}^\infty p_k(x,\nu )\la^k=(1+\la )^\nu \exp (x\la )
$$ 
and $p_{k}=0$ for $k<0$. 
\end{Thm}  

\noindent {\bf Remarks.} The polynomials $p_{k}((x,\nu )$ are 
essentially just associated Laguerre polynomials, 
and, for each $n$, $\mathcal{D}_n(x,\nu )$ is a Schur polynomial with restricted arguments, 
corresponding to the partition $(n,n-1,\ldots ,2,1)$.  
The result stated above is an adapted form of Theorem 1 in \cite{KM}, and 
describes one family of rational solutions of $P_{III}$; for a complete 
description of all rational solutions of $P_{III}$ for $\gamma\delta\neq 0$, see \cite{kaji}. 
The polynomials $\mathcal{D}_n(x,\nu )$ (after some scaling) are 
known as the Umemura polynomials for $P_{III}$. 
Further properties of scaled Umemura polynomials for $P_{III}$ are 
detailed in \cite{peterp3}, including the remarkable patterns formed by the roots, 
and differential/difference equations; analogous polynomials 
corresponding to the special cases when $\gamma\delta=0$ are also treated 
there.   

\begin{Thm} \label{jacobitrudi}  
The discrete Yablonskii-Vorob'ev polynomials are given in terms of 
determinants of 
Jacobi-Trudi type by the formula  
\beq\label{yabum} 
Y_n (t,h) = c_n \, h^{n(n+1)/2}\, 
\mathcal{D}_n\left(\frac{4}{h^3},\frac{t}{h}-\frac{4}{h^3}\right), 
\eeq 
where 
\beq \label{cform}  
c_n =(2n-1)!!(2n-3)!!\ldots 3!! 1!! 
\eeq 
for $n\in\mathbb{N}$. 
\end{Thm}

The proof of the preceding theorem makes use of some results in the next section, 
and is relegated to the appendix.
However, it is clear that if we rearrange the formula (\ref{yabum}) 
then we can rewrite $\mathcal{D}_n(x,\nu )$ in the form of a Hankel determinant. 

\begin{Cor} 
The Umemura polynomials for $P_{III}$, given in scaled form by 
$\mathcal{D}_n(x,\nu )$, are proportional to the Hankel determinants $u_n$ as in   
(\ref{eq:dhank2}) with 
$h=(x/4)^{-1/3}$ and $t=(x/4)^{-1/3}(x+\nu )$ . 
\end{Cor} 

\noindent {\bf Remark.} It is known that the 
Painlev\'e differential equations form a coalescence cascade from 
$P_{VI}$ down to $P_I$ (see \cite{ince}). 
In \cite{KM} it is shown that the coalescence limit from $P_{III}$ to 
$P_{II}$ produces the Yablonskii-Vorob'ev polynomials $y_n(t)$ as 
a limit of the Umemura polynomials, but this arises in a different way compared 
with the limit $h\to 0$ considered above. More precisely, with the scaling used here, 
the coalescence from (\ref{p3}) to (\ref{eq:p2a}) arises when the independent 
variable $x$ and parameter $\nu$ scale as 
$ 
x=\frac{t}{\epsilon}+\frac{4}{\epsilon^3}
$, 
$\nu = \frac{1}{2}- \frac{4}{\epsilon^3}$, 
with $\epsilon\to 0$. In this limit, up to scaling  the polynomials 
$\mathcal{D}_n(t/\epsilon +4/\epsilon^3,1/2- 4/\epsilon^3)$ produce 
$y_n(t)$ at leading order in $\epsilon$. 

\section{Symplectic properties and discrete $P_{XXXIV}$}

Okamoto \cite{okamoto} showed that $P_{II}$ can be written as the system 
(\ref{p2sys}) 
which is in Hamiltonian form, i.e. 
\beq
\dot{q}=\frac{\partial H}{\partial p}, \qquad  
\dot{p}=-\frac{\partial H}{\partial q},
\label{eq:ham1}
\eeq
with
\beq
H=\frac{p^2}{2}-(q^2+\frac{t}{2})p-\ell q.
\label{eq:ham2}
\eeq
Eliminating $p$ gives (\ref{eq:p2a}), 
which is $P_{II}$, whilst eliminating $q$ gives 
\beq
\ddot{p}=\frac{\dot{p}^2}{2p}+2p^2-tp-\frac{\ell^2}{2p}
\label{eq:ham3}
\eeq
which is known as $P_{XXXIV}$ (see \cite{ince}, chapter XIV).  

This representation has been used in \cite{pet1} to study further properties of the Yablonskii-Vorob'ev
polnomials. Although the $alt$-$dP_{II}$ equation (\ref{eq:dp2a}), being a non-autonomous difference 
equation, does not have a Hamiltonian form, many of the results proved there have 
their counterparts in the discrete case.  
For ease of comparison, we briefly recall some known results on $P_{II}$. 
In terms of the canonical coordinates 
$(q_{n},p_{n})$, 
the B\"acklund
transformation for $P_{II}$ can be written (together with its inverse) as 
\beq 
\fl \begin{array}{l}    
q_{n+1}=-q_{n}-\frac{\ell }{p_{n}}, \qquad   
p_{n+1}=-p_{n}+t+2\left(q_{n}+\frac{\ell}{p_{n}}\right)^2 ; \\
q_{n-1}=-q_{n}-\frac{(\ell -1)}  {2q_{n}^2-p_{n}+t}, \quad  
p_{n-1}=-p_{n}+t+2q_{n}^2.  
\end{array}
\label{eq:hbac1}
\eeq
It is straightforward to check that 
$dq_{n+1} \wedge dp_{n+1}+dH_{n+1}\wedge dt =dq_{n} \wedge dp_{n} + dH_{n}\wedge dt$, 
so the transformation $n\rightarrow n+1$  is a canonical transformation 
on the extended phase space with coordinates $(q_{n},p_{n},t)$.     
The generating function for this canonical transformation is
\beq
\mathcal{F} (q_{n},q_{n+1})=\ell\log(q_{n+1}+q_{n})+\frac{2}{3}q_{n+1}^3+tq_{n+1}
\label{eq:hbac}
\eeq
so that 
\[
p_{n}=-\frac{\partial \mathcal{F}}{\partial q_{n}}, \qquad 
p_{n+1}=\frac{\partial \mathcal{F}}{\partial q_{n+1}}.  
\]
The B\"acklund
transformation formulae (\ref{eq:hbac1}) imply that 
any sequence of solutions $q_n$ of $P_{II}$ (labelled by $n=\ell -1/2$) 
satisfies  
\beq
\frac{\ell}{q_{n+1}+q_{n}}+\frac{\ell-1}{q_{n}+q_{n-1}}+2q_{n}^2+t=0,
\label{eq:ham6}
\eeq
which is a discrete form of $P_{I}$, whilst $p_{n}$ satisfies
\beq 
\fl  
(p_{n+1}-p_{n-1})^2p_{n}^4-4\ell^2(p_{n+1}+2p_{n}+p_{n-1}-2t)p_{n}^2+4\ell^4=0.
\label{eq:ham7}
\eeq

We will now show how the above results carry over into the discrete case.   
\begin{Pro} 
In terms of the variables $q_n=(g_n+1)/h$ and $p_n$, the B\"acklund
transformation (\ref{gbt}) 
for the $alt$-$dP_{II}$ equation (\ref{eq:dp2a}) (corresponding to $n\to n+1$) 
can be written together with its inverse 
(corresponding to $n\to n-1$) as    
\beq
\fl 
\begin{array}{rcl}  
q_{n+1}&=&\frac{q_{n}p_{n}+\ell}{h(q_{n}p_{n}+\ell )-p_{n}},  \\
p_{n+1}&=&(q_{n}p_{n}+\ell )^2\left(\frac{2}{p_{n}^2} -\frac{h^2}{p_{n}}\right) 
-\frac{h}{p_{n}}(t-2p_{n})(q_{n}p_{n}+\ell ) -p_{n}+t; \\
q_{n-1}&=&-q_{n}+\frac{1-\ell +hq_{n}(-2q_{n}^3+q_{n}p_{n}-q_{n}t+\ell -1)+h^{2}q_{n}^{3}(-2p_{n}+t)+h^3q_{n}^4p_{n}}
{-p_{n}+t+2q_{n}^3-hq_{n}(2q_{n}^2-3p_{n}+2t) 
+h^2q_{n}^2(-3p_{n}+t)+h^3q_{n}^3p_{n}}, \\
p_{n-1}&=&-p_{n}+t+2q_{n}^2+h(2p_{n}-t)q_{n}-h^2p_{n}q_{n}^2.
\label{eq:hbac5}
\end{array}  
\eeq 
This B\"acklund
transformation is a symplectic map in the phase space 
with coordinates $(q_n,p_n)$, with generating function 
\bea
\fl
\mathcal{F}(q_{n},q_{n+1})&=&\frac{2q_{n+1}}{h^2}+\frac{2}{h^3(1-hq_{n+1})}-\frac{1}{h^3}\Big(h^2t-4+\ell h^3\Big)\log(1-hq_{n+1})\nonumber \\
&&+\ell\log(hq_{n}q_{n+1}-q_{n+1}-q_{n}) 
\label{eq:symp3}
\eea
such that $d\mathcal{F}=p_{n+1}\, dq_{n+1}-p_{n}\, dq_{n}$; 
in other words the canonical Poisson bracket 
$\{ q_n,p_n \}=1$ is preserved for $n\to n+1$. 
The equations (\ref{eq:hbac5}) and the generating function reduce to 
those of the B\"acklund
transformation for $P_{II}$ in the continuum limit $h\to 0$.   
\end{Pro} 
\begin{prf} 
The formulae (\ref{eq:hbac5}) follow immediately from  Lemma \ref{dmiura} 
and Corollary \ref{miuracor}. To verify that the map 
$(q_n,p_n)\mapsto (q_{n+1},p_{n+1})$ is symplectic, it is sufficient to calculate  
directly that its 
Jacobian determinant is equal to 1. 
This also follows directly from the closure of the exact one-form $d\mathcal{F}$, 
upon checking that   
$p_n=-\frac{\partial \mathcal{F}}{\partial q_n}$, 
$p_{n+1}=\frac{\partial \mathcal{F}}{\partial q_{n+1}}$ with 
the generating function $\mathcal{F}$ as in the formula (\ref{eq:symp3}). 
It is  straightforward to verify that the relations (\ref{eq:hbac5}) have the correct 
continuum limit given by (\ref{eq:hbac1}), and that (\ref{eq:symp3}) also 
yields  (\ref{eq:hbac}) when $h\to 0$. 
$\,\,\Box$ 
\end{prf} 
\noindent {\bf Remark.} 
The symplectic structure for the $alt$-$dP_{II}$
equation can be derived from Okamoto's Hamiltonian formulation 
for $P_{III}$, since the canonical coordinates $(q_n,p_n)$ 
above are related to those for $P_{III}$ by a shift and rescaling 
(cf. section 2.3 in \cite{peterp3} for instance).   
\begin{Cor} \label{p3rd} 
For the 
$alt$-$dP_{II}$ 
equation (\ref{eq:dp2a}), 
the 
analogue of  (\ref{eq:ham6}) is 
the equation 
\beq
\fl
\frac{\ell }{q_{n+1} +q_{n}-hq_{n+1}q_{n}} + 
\frac{(\ell -1)}{q_{n}+q_{n-1}-hq_{n}q_{n-1}}
+\frac{2q_{n}^{2}}{1-hq_n} + t-h\ell 
=0.
\label{eq:hbac6}
\eeq
In terms of $g_n$ this is the second order difference equation 
\beq \fl 
\label{altdp2n} 
\frac{\ell}{g_{n+1}g_n-1} 
+ 
\frac{(\ell -1)}{g_ng_{n-1}-1} 
+\frac{2}{h^3}\left(g_n +\frac{1}{g_n}\right) 
+\ell +\frac{4}{h^3}-\frac{t}{h}=0,  
\eeq 
which is another form of the $alt$-$dP_{II}$ equation. 
The conjugate momentum $p_n$ 
satisfies a third order recurrence relation in $n$, namely 
\beq 
\fl 
\begin{array}{l} 
p_{n+1}\{\frac{p_{n}}{2n+1}-\frac{hp_{n}[(2n-1)^2+2p_{n-1}^2(p_{n}-p_{n-2})]}{4p_{n-1}(4n^2-1)}-
\frac{h^2p_{n}t}{4(2n+1)}+\frac{h^3p_{n}(2n-1)}{8(2n+1)}\} 
\\
+p_{n-2}\{-\frac{p_{n-1}}{2n-1}-\frac{hp_{n-1}[(2n+1)^2+2p_{n}^2(p_{n-1}-p_{n+1})]}{4p_{n}(4n^2-1)}+
\frac{h^2p_{n-1}t}{4(2n-1)}+\frac{h^3p_{n-1}(2n+1)}{8(2n-1)}\}
 \\
+\frac{h^2p_{n}^2p_{n-1}^2}{2(4n^2-1)}+\frac{h(4n^2-1)}{8p_{n}p_{n-1}}
-\frac{(2n+1)[h^3(2n-1)-2h^2t+8]}{16p_{n}}
 \\
-\frac{(2n-1)[h^3(2n+1)+2h^2t-8]}{16p_{n-1}}
+p_{n}p_{n-1}[\frac{8-2h^2t+(4n^2-3)h^3}{4(4n^2-1)}]-\frac{(6n+5)hp_{n}}{4(2n+1)} \\
-\frac{(6n-5)hp_{n-1}}{4(2n-1)} +\frac{h}{32}[32t+16h-4h^2t^2-4h^3t+(4n^2-1)h^4]=0.
\end{array}
\label{eq:hbac8}
\eeq 
\end{Cor} 
\begin{prf} 
Upon solving the first of (\ref{eq:hbac5}) for $p_{n}$ and substituting in the last one, 
(\ref{eq:hbac6}) results. 
One can eliminate the quadratic terms in $q_{n}$ from the second and fourth of (\ref{eq:hbac5}) to get
\beq 
q_{n}=-\frac{2n+1}{4p_{n}}-\frac{2(p_{n+1}-p_{n-1})p_{n}+(t-2p_{n})(2n+1)h}{2(2n+1)(h^2p_{n}-2)}
\label{eq:hbac7}
\eeq  
After substituting this into the first of (\ref{eq:hbac5}) 
with $n\rightarrow n-1$, we obtain (\ref{eq:hbac8}). 
\end{prf} 
\noindent {\bf Remarks.} The fact that the $alt$-$dP_{II}$ equation 
is self-dual, in the sense that the superposition formula (\ref{altdp2n}) 
for its 
B\"ackund transformation is (up to rescaling and reversing the roles 
of the dependent variable and the B\"acklund parameter) the equation 
itself, was first noted in \cite{FSKGR}.  
Equation (37) in \cite{pet1} 
is the continuum limit of equation (\ref{eq:hbac8}).  
The latter relation allows one to obtain $p_{n+1}$ 
uniquely given $p_{n},p_{n-1},p_{n-2}$. Remembering that
$p_{n}=\frac{\ta_{n-1}\ta_{n+1}}{2\ta_{n}^2}$, we see that if $\ta_j(t,h)$ for $j=n-3,n-2,\ldots , n+1$ 
are given for a particular value of $t$ , 
then $\ta_{n+2}$ can be evaluated for this same value of $t$. More precisely, (\ref{eq:hbac8}) is equivalent 
to a recurrence relation for the $\ta_{n}$ 
that involves no shifts in $t$ and is linear in $\ta_{n+2}$ and $\ta_{n-3}$; in particular this relation, 
which is of fifth order in $n$,  
is satisfied by the discrete Yablonskii-Vorob'ev polynomials $Y_n$ for $n\in\mathbb{Z}$. 
The latter recurrence for the tau-functions, which is omitted here, 
tends to equation (38) in \cite{pet1} in the continuum limit.  
It is also possible to write $q_{n}$ in terms of unshifted $\ta_{n}$,
by substituting the right hand side of (\ref{ptau}) for $p_{n}$ in (\ref{eq:hbac7}).

We now consider the map in the $(q_n,p_n)$ phase space corresponding to 
shifting  
in $t$ rather than $n$. 
\begin{Pro} 
In the phase space with coordinates $(q_n,p_n)$, 
the shift $t\to t+h$ corresponding to the $alt$-$dP_{II}$ system 
(\ref{dp2sys}) is given by 
\beq
\fl 
\begin{array}{rcl}
\ovl{q}_{n}& = & \frac{-2q_{n}+h(t-2p_{n})+h^2p_{n}q_{n}}{-2-2hq_{n}+h^2(t-p_{n})+h^3q_{n}p_{n}}, \\ 
\ovl{p}_{n} & = & \frac{1}{(h^2p_{n}-2)^2}\Big(-2-2hq_{n}+h^2(t-p_{n})+h^3q_{n}p_{n}\Big)\times \\
&& \times \Big(-2p_{n}-2h(\ell +q_{n}p_{n})+ h^2p_{n}(t-p_{n})+h^3p_{n}(\ell +q_{n}p_{n})\Big). 
\end{array} 
\label{eq:symp4}
\eeq
This is a symplectic map with the generating function 
\beq
\fl
\begin{array}{rcl} 
S&=&\frac{1}{h^3}(4-h^2t)\log(2-h^2p_{n})+\frac{\ell }{2}\log\frac{\ovl{p}_{n}}{p_{n}}+\frac{1}{h}\Big(p_n-\ovl{p}_n\Big)   \\
 &+&\frac{1}{h}  \sqrt{\ell^2 h^2+4p_{n}\ovl{p}_{n}}-\ell \tanh^{-1}\left(\frac{\sqrt{\ell^2h^2 
+4p_{n}\ovl{p}_{n}}}{\ell h}\right)
\end{array} 
\label{eq:symp8}
\eeq
such that $dS=q_n\, dp_n -\ovl{q}_n\, d\ovl{p}_n $.
\end{Pro} 
\begin{prf}
The equations (\ref{eq:symp4}) just correspond to rewriting (\ref{dp2sys}) in terms 
of $q_n=(g_n+1)/h$, $\ovl{q}_n=(\ovl{g}_n+1)/h$. It is extremely easy to check directly 
from (\ref{dp2sys}) 
that the symplectic form $\omega_n = dq_n\wedge dp_n = \frac{1}{h}dg_n\wedge dp_n$ is preserved by 
the shift in $t$. To find the generating function, it is convenient to
write $g_{n},\ovl{g}_{n}$ in terms of $p_{n},\ovl{p}_{n}$, giving 
\beq
\fl
\begin{array}{rcl}
\frac{\partial \hat{S}}{\partial p_{n}}&=&g_{n}=-\left(\frac{4-h^2t}{2-h^2p_{n}}\right) -
\frac{1}{2p_{n}}\Big( \ell h\mp \sqrt{\ell^2h^2+4\ovl{p}_{n}p_{n}}\Big),  \\
\frac{\partial \hat{S}}{\partial \ovl{p}_{n}}&=&-\ovl{g}_{n}=\frac{1}{2\ovl{p}_{n}}\Big(\ell h\pm 
\sqrt{\ell^2h^2+4\ovl{p}_{n}p_{n}}\Big). 
\end{array} 
\label{eq:symp7}
\eeq 
(One has to take the upper choice of sign in each case to get the correct continuum limit.)  
Having found an $\hat{S}$ for which the relations (\ref{eq:symp7}) 
hold,   $S=(\hat{S}+p_{n}-\ovl{p}_{n})/h$ provides the generating function in (\ref{eq:symp8}). 
$\,\,\Box$  
\end{prf} 
\begin{Cor} 
The discrete $P_{XXXIV}$ equation associated with the $alt$-$dP_{II}$ equation (\ref{eq:dp2a}) can 
be written either as 
\beq
\fl 
\pm \sqrt{\ell^2h^2+4p_{n}\ovl{p}_{n}} \pm \sqrt{\ell^2h^2+4p_{n}\unl{p}_{n}} 
=\frac{2p_{n}(4-h^2t)}{2-h^2p_{n}},
\label{eq:p34a}
\eeq
or with the square root signs removed as 
\[
\fl
\Big(\ell^2h^2+4p_{n}\ovl{p}_{n}\Big) 
\Big(\ell^2h^2+4p_{n}\unl{p}_{n}\Big) 
=\left( \frac{2p_{n}^2(4-h^2t)^2}{(2-h^2p_{n})^2} 
-\ell^2h^2-2p_{n}(\ovl{p}_{n}+\unl{p}_{n})\right)^2.
\]
\end{Cor} 
\begin{prf}
The second order recurrence relation for $p_{n}$ is obtained by downshifting the second of (\ref{eq:symp7}) and equating
it to minus the first. 
$\,\,\Box$ 
\end{prf} 
\noindent{\bf Remark.} The 3-point correspondence (\ref{eq:p34a}) 
is equivalent to an analogous equation for the variable 
$y_n$, that is equation (5.3) in \cite{FSKGR}, and the equation (\ref{eq:p34a}) 
has the same structure as certain discrete Ermakov-Pinney equations 
constructed in \cite{cmus}. The presence of this structure is due to the 
connection with discrete Schr\"odinger equations (for which, see the proof 
of Proposition \ref{dlax} below).

\section{Lax pair for $alt$-$dP_{II}$}

The continuum $PII$ is equivalent to the pair of bilinear equations (\ref{eq:blb}). A Lax pair 
for $P_{II}$  is given by the linear problem 
\beq
\frac{\partial \Psi}{\partial t}=\mathcal{B}\Psi, \qquad  \frac{\partial \Psi}{\partial \eta}=\mathcal{A}\Psi,
\label{eq:mon1}\eeq
where
\bea
\mathcal{B}=\left(\begin{array}{cc}
-\frac{\eta}{2}&-\frac{i\tau_{n+1}}{2\tau_{n}}\\
\frac{i\tau_{n-1} }{2\tau_{n}}& \frac{\eta}{2}
\end{array} \right)
\label{eq:mon2}
\eea
and
\bea
\mathcal{A}=\left(\begin{array}{cc}
-\eta^2-\frac{t}{2}+\frac{\tau_{n-1}\tau_{n+1}}{2\tau_{n}^2}&
-\frac{i\eta\tau_{n+1}}{\tau_{n}(t)}+i\frac{d}{dt}\left(\frac{\tau_{n+1}}{\tau_{n}}\right)\\
 \frac{i\eta\tau_{n-1}}{\tau_{n}}+i\frac{d}{dt}\left(\frac{\tau_{n-1}}{\tau_{n}}\right)&
+\eta^2+\frac{t}{2}-\frac{\tau_{n-1}\tau_{n+1}}{2\tau_{n}^2}
\end{array}\right). 
\label{eq:mon3}
\eea
Consistency of (\ref{eq:mon1}) requires
\beq
\frac{\partial \mathcal{B}}{\partial \eta}-\frac{\partial \mathcal{A}}{\partial t}+[\mathcal{B},\mathcal{A}]=0,
\label{eq:mon5}
\eeq
leading to the two conditions
\beq\label{schro} 
\ddot{\phi}_{\pm}+V\phi_{\pm}=0, 
\eeq 
where we have set 
\[ 
\phi_{\pm}=\mp \frac{i\ta_{n\pm 1}}{2\ta_n} 
, \qquad V=\frac{t}{2}-2\phi_+ \phi_-. 
\]
The choice of gauge $V=t/2-2\phi_+ \phi_- =2\frac{d^2}{dt^2}\log\tau_n$ 
gives precisely the equation (\ref{eq:bld}), and with this normalisation for 
the tau-functions  
the conditions (\ref{schro}) give the first equation in (\ref{eq:blb}) f 
or $F=0$, 
together with the same equation for $n\to n+1$. The bilinear equation (\ref{eq:blc}) 
is a consequence, and the second equation in (\ref{eq:blb}) 
then follows. It is well known that (\ref{eq:mon5}) is an 
isomonodromy condition: the monodromy of the solutions 
of the second linear equation (\ref{eq:mon1}) in the complex $\eta$ plane 
is independent of $t$ if and only if $P_{II}$ holds.  

In the discrete case the situation is completely analogous, based on a linear problem 
that comes from the first part of the discrete Toda Lax pair (\ref{eq:dlax1}). 
\begin{Pro}\label{dlax}   
The linear problem   
\beq
\overline{\Psi}=B\Psi, \qquad \lambda\partial_{\lambda}\Psi=A\Psi 
\label{eq:dmon1}
\eeq
where
\bea
B=\left(\begin{array}{cc}
\frac{1}{\lambda}&\frac{h\tau_{n+1}}{2i\tau_{n}}\\
-\frac{h\tau_{n-1}}{2i\tau_{n}}&\lambda
\end{array}\right)=\left(\begin{array}{cc}
\frac{1}{\lambda}&\phi_+\\
\phi_- &\lambda
\end{array}\right)
\label{eq:dmon2}
\eea
and
\[ \fl 
A=\left(\begin{array}{cc}
\frac{2}{h^3}\left(-\frac{1}{\lambda^2}+2(\phi_+\unl{\phi}_-+1)-\lambda^2 \right) -\frac{t}{h}+\frac{1}{2} & 
\frac{4}{h^3}\left(-\frac{\phi_+}{\lambda} +\unl{\phi}_+\lambda \right) \\ 
 \frac{4}{h^3}\left(-\frac{\unl{\phi}_-}{\lambda} +{\phi}_-\lambda \right)  & 
\frac{2}{h^3}\left(\frac{1}{\lambda^2}-2(\unl{\phi}_+\phi_-+1)+\lambda^2 \right) +\frac{t}{h}-\frac{1}{2}
\end{array}\right) 
\label{eq:dmon3}
\]
constitutes a Lax pair for the $alt$-$dP_{II}$ equation (\ref{eq:dp2a}), 
which is equivalent to the consistency 
condition 
\beq
\lambda\partial_{\lambda}B+BA-\overline{A}B=0
\label{eq:dmon5}
\eeq 
for (\ref{eq:dmon1}). In the limit $h\to 0$, this linear system reduces to the Lax pair (\ref{eq:mon1}) for $P_{II}$.  
\end{Pro} 
\begin{prf}
The consistency condition (\ref{eq:dmon5}) implies the two relations 
\beq\label{dschro} 
\ovl{\phi}_{\pm}+V\phi_\pm +\unl{\phi}_{\pm}=0,
\eeq 
which take the form of discrete Schr\"odinger equations 
with 
\[\fl 
V=\frac{(h^2t-4)}{2(1-\phi_+\phi_-)}=\frac{2\ta_n^2(h^2t-4)}{4\ta_n^2-h^2\ta_{n+1}\ta_{n-1}}. 
\] 
If we fix the gauge so that $V=-2\ta_n^2/(\ovl{\ta}_n\unl{\ta}_n)$, then we get 
precisely the relation (\ref{eq:qd}), and upon substituting the latter 
expression for $V$ into each of the equations (\ref{dschro}) in turn we 
find that (\ref{eq:qb}) and (\ref{eq:qbbt}) hold. Then by 
Theorem \ref{normalise}, $g_n =-\unl{\ta}_{n-1}\ta_n/(\ta_{n-1}\unl{\ta}_n)$ 
satisfies the  $alt$-$dP_{II}$ equation. For the continuum limit 
one should take $\lambda=e^{\frac{h\eta}{2}}$, which gives 
\beq
A=\frac{2}{h}\mathcal{A}+\mathcal{O}(h^{0}),B=1+h\mathcal{B}+\mathcal{O}(h^{2}), 
\label{eq:dmon8}
\eeq
so that the condition (\ref{eq:mon5}) arises from (\ref{eq:dmon5}) as $h\rightarrow 0$.
$\,\,\Box$ 
\end{prf} 

\noindent {\bf Remark.} In \cite{FSKGR} a different $2\times 2$ 
Lax pair is presented for the $alt$-$dP_{II}$ equation, by reduction from 
the modified Boussinesq lattice. However, we have not found 
a direct relationship between these two Lax pairs. 

\section{Concluding remarks} 
There are many ways to construct a discretisation of a given integrable differential 
equation, depending on which particular properties (e.g. Lax pair, explicit solutions, 
Poisson structure, Hirota bilinear form,$\ldots$) one most wishes to preserve. 
(For a thorough account of the Hamiltonian approach, see \cite{surisbook}.)
Due to the non-uniqueness of discretisation, it is not always clear what is the 
``best'' discrete analogue of a continuous system. The derivation of the 
$alt$-$dP_{II}$ equation presented here was 
initially motivated by the construction of exact rational solutions, but it has turned 
out that analogues of all the other structures associated with the second  
Painlev\'e equation arise naturally here as well. The fact that this discretisation scheme 
for  $P_{II}$, based on the discrete Toda lattice, turned out to be connected with the superposition formula for 
$P_{III}$ was completely unexpected by us, but led to different expressions for the polynomial
tau-functions in terms of Jacobi-Trudi determinants. In \cite{FSKGR} other solutions 
of $alt$-$dP_{II}$ were constructed in terms of Casorati determinants of discrete 
Airy functions. 
In future we should like to analyse 
other solutions of the equation (\ref{eq:dp2a}). We constructed the rational solutions 
of this equation from polynomial tau-functions given by Hankel determinants, but 
recent results for the continuous case \cite{josh} lead us to expect 
that all tau-functions should have a similar structure.

\ack 
We are grateful to Peter Clarkson for interesting conversations on related matters. 
We would like to thank all the referees for their 
instructive comments, in particular for helping us 
to identify   
the links  between the discrete Yablonskii-Vorob'ev polynomials, 
$alt$-$P_{II}$,   
$P_{III}$ and the Umemura polynomials, and for pointing out important 
references including 
\cite{FSKGR}.

\appendix 

\section*{Appendix} 

\setcounter{section}{1} 

Here we present the proof of Theorem \ref{jacobitrudi}. The result essentially 
follows from 
the fact that the B\"acklund transformation for $P_{III}$ generates a 
sequence of rational solutions $w_n$ given by (\ref{wform}), which  
simultaneously provide rational solutions of the $alt$-$dP_{II}$ equation (\ref{adp2s})  
by setting $g_n=-1/w_n$ with $z_n=n+1/2$ for  $n\in\mathbb{Z}$ and $\mu=-\nu$. On the   
other hand, Theorem \ref{ratthm} and Corollary \ref{p3rd} together imply that 
the $alt$-$dP_{II}$ equation in the form (\ref{altdp2n}) has rational solutions, given 
by suitable ratios of discrete Yablonskii-Vorob'ev polynomials, when $\ell = n+1/2$  
with $n\in\mathbb{Z}$. Upon comparing (\ref{adp2s}) with (\ref{altdp2n}) we see that these 
two sets of rational solutions coincide if we identify $x=4/h^3$, $\nu = t/h -4/h^3$, 
and then it follows from Theorem \ref{kajmas} and Theorem \ref{ratthm} that 
\beq \label{relndet} 
w_n = \frac{Z_{n-1}(t,h)Z_n(t-h,h)}{Z_{n-1}(t-h,h)Z_n(t,h)} = 
\frac{Y_{n-1}(t,h)Y_n(t-h,h)}{Y_{n-1}(t-h,h)Y_n(t,h)} 
\eeq 
for all $n\in\mathbb{Z}$, where $Z_n(t,h)$ denotes the right hand side of (\ref{yabum}) for $n\geq 0$, and 
the identity extends to negative $n$ upon setting $Z_{-n}=Z_{n-1}$. Then 
it is necessary to show that $Y_n(t,h)=Z_n(t,h)$ for all $n$. 
It is sufficient to consider $n\in\mathbb{N}$ as in Theorem \ref{jacobitrudi}, as the extension 
to negative $n$ is trivial. 

For $n=0$ the result $Y_0=1=Z_0$ is obvious, so to use induction 
we assume that $Y_{N-1}=Z_{N-1}$ and then from (\ref{relndet}) with $n=N$ it holds that 
\beq\label{ratios} 
\frac{Z_{N}(t-h,h)}{Z_{N}(t,h)} = 
\frac{Y_{N}(t-h,h)}{Y_{N}(t,h)}. 
\eeq
Both sides of the relation (\ref{ratios}) are ratios of polynomials in $t$ (for $Z_{N}$ this follows 
from the determinant formula for $\mathcal{D}_{N}$), and the rational 
functions on each side must have the same zeroes and poles. However, although Theorem \ref{zerothm} 
implies that the numerator and denominator on the right hand side have no common factors, we 
cannot immediately assert that the same is true on the left hand side without knowing the degree 
of $\mathcal{D}_N(x,\nu)$ in $\nu$. 
(The proof of Corollary 2 in \cite{KM} just gives the degree in $x$ of $\mathcal{D}_N$, 
denoted $F_N$ there,  as $N(N+1)/2$.)  
However, by Proposition 3 in \cite{KM} these Jacobi-Trudi determinants satisfy the 
recurrence  
\beq\label{drec} 
\fl 
(2n+1)\mathcal{D}_{n+1}\mathcal{D}_{n-1}+ 
x(\mathcal{D}_n\mathcal{D}_n''-(\mathcal{D}_n')^2)+\mathcal{D}_n\mathcal{D}_n' 
-(x+\nu )\mathcal{D}_n^2=0, 
\eeq 
with $\mathcal{D}_{-1}=1=\mathcal{D}_0$ and $'$ denoting $d/dx$. In fact, this recurrence 
can be used to show that $\mathcal{D}_N(x,\nu )$ is also of degree $N(N+1)/2$ in $\nu$, 
but we do not need this. 
Instead, setting $\nu =0$ in (\ref{drec}) 
leads to the expression $\mathcal{D}_n(x,0)=c_n^{-1}x^{n(n+1)/2}$ where $c_n$ is given in terms of 
double factorials by (\ref{cform}).  Hence for the case at hand we have 
$Z_N(4/h^2,h)=c_n h^{n(n+1)/2}\mathcal{D}_n(4/h^3,0)=(4/h^2)^{n(n+1)/2}=Y_N(4/h^2,h)$ 
by Lemma \ref{eq:zer1}. Using (\ref{ratios}) it then follows by induction that 
$Z_N(4/h^2 +kh,h)=Y_N(4/h^2+kh,h)$ for all $k\in\mathbb{Z}$, and since these 
two polynomials agree for infinitely many values of $t$ they must be equal, as required. $\Box$

\end{document}